\documentstyle[11pt,aaspp4]{article}
\received{}
\accepted{}
\input tex.def

\slugcomment{To appear in {\it The Astrophysical Journal}}

\begin{document}

\title{The Influence of Bars on Nuclear Activity}

\author{Luis C. Ho}
\affil{Department of Astronomy, University of California, Berkeley, CA
94720-3411}

\and

\affil{Harvard-Smithsonian Center for Astrophysics, 60 Garden St., Cambridge,
MA 02138\footnote{Present address.}}

\author{Alexei V. Filippenko}
\affil{Department of Astronomy, University of California, Berkeley, CA
94720-3411}

\and

\author{Wallace L. W. Sargent}
\affil{Palomar Observatory, 105-24 Caltech, Pasadena, CA 91125}
 
\begin{abstract}
 
Gravitational torques induced by a stellar bar on the interstellar medium of 
a disk galaxy instigate radial inflow of gas toward the central regions of 
the galaxy.  Accordingly, the presence of a bar should reinforce nuclear star 
formation activity or the fueling of an active galactic nucleus (AGN).  We 
test this hypothesis by comparing the detection rate and intensity of nuclear 
\hii\ regions and AGNs among barred and unbarred galaxies in a sample 
of over 300 spirals selected from our recent optical spectroscopic survey of 
nearby galaxies.  The AGN group includes Seyfert nuclei as well as 
low-ionization nuclear emission-line regions (LINERs).  

Among late-type spirals (Sc--Sm), but not early-type (S0/a--Sbc), we observe
in the barred group a very marginal increase in the detection rate of \hii\ 
nuclei and a corresponding decrease in the incidence of AGNs.  The minor 
differences in the detection rates, however, are statistically insignificant, 
most likely stemming from selection effects and not from a genuine influence 
from the bar.  The presence of a bar seems to have no noticeable impact on the 
likelihood of a galaxy to host either nuclear star formation or an AGN.  

The nuclei of early-type (S0/a--Sbc) barred spirals do exhibit measurably 
higher star-formation rates than their unbarred counterparts, as indicated
by either the luminosity or the equivalent width of H\al\ emission.  By 
contrast, late-type spirals do not show such an effect.  These results agree 
with previous studies and can be explained most easily in terms of 
structural differences between bars in early-type and late-type spirals.  
Nuclear \hii\ regions spanning a wide range of intensity are found regardless 
of the presence of a bar, suggesting that a bar is neither a necessary nor 
a sufficient condition for star formation to occur in galactic nuclei.  Other 
factors, such as the availability of gas, must be equally important.

Bars, on the other hand, have a negligible effect on the strength of 
the AGNs in our sample, regardless of the Hubble type of the host 
galaxy.  This result confirms similar conclusions reached by other studies 
based on much smaller samples.  Assuming that AGNs are fueled by gas from 
the interstellar medium of the host galaxy, some inferences concerning the 
fueling process can be made.  We speculate that inner Lindblad 
resonances, particularly common in barred galaxies with large 
bulge-to-disk ratios, prevent gas which has been radially transported from 
large scales from reaching the nucleus.  We discuss the feasibility of 
sustaining the power output of nearby AGNs with debris from tidal disruption 
of stars by a supermassive black hole, as well as with mass loss from evolved 
stars, and conclude that such processes should be sufficient to fuel the 
low-luminosity nuclear sources found in many nearby galaxies.  

\end{abstract}


\section{Introduction}

The search for environmental factors which might influence activity in 
galaxies has had a long observational and theoretical history (see the 
many contributions in Sulentic, Keel, \& Telesco 1990 and in Shlosman 1994).
Much of the effort has focused on the role of galaxy-galaxy interactions 
in generating activity in galactic nuclei, of which two types commonly are 
discussed --- starburst (\hii) nuclei and ``nonstellar'' active galactic 
nuclei (AGNs).  
Numerical simulations indicate that gas flows toward the centers of galaxies 
in response to tidal perturbations and the resulting dissipation of angular 
momentum (see Barnes \& Hernquist 1992, and references therein).  This
theoretical prediction qualitatively explains a number of observational 
findings.  Relative to control samples of isolated galaxies, there have been 
reports of an increased frequency and strength of nuclear star formation in 
interacting or paired galaxies, a statistically significant excess of 
companions associated with Seyfert galaxies and low-redshift QSOs, and a 
higher incidence of AGN and starburst host galaxies with disturbed 
morphologies (Phinney 1994; Heckman 1994; and references therein).

Besides tidal interactions, the presence of other non-axisymmetric 
gravitational perturbations, such as those produced by a large-scale stellar 
bar or an oval distortion, can also induce radial gas inflow.  Such a mechanism 
for fueling nuclear activity has been discussed by Simkin, Su, \& Schwarz 
(1980), Shlosman, Frank, \& Begelman (1989), Friedli \& Benz (1993),  Heller 
\& Shlosman (1994), among many others.  De Jong \etal (1984) noticed that 
optically-selected barred spirals tend to have higher infrared (IR) 
luminosities and hotter 100 to 60 \micron\ colors than their unbarred 
counterparts.  Similarly, Hawarden \etal (1986) discovered that more than 
one-third of barred spirals emit excess radiation at 25 \micron\ relative to 
unbarred spirals, and although the coarse {\it IRAS} beam could not reveal 
the location of the emission within each galaxy, they postulated that the 
excess emanates from enhanced circumnuclear star formation.  Puxley, Hawarden, 
\& Mountain (1988) confirmed through radio observations that the emission 
is indeed confined mainly to the central 1--3 kpc, and Devereux (1987, 1989) 
demonstrated that star formation powers most of the activity.  Hummel \etal 
(1990) reached similar conclusions, emphasizing that the enhanced radio 
emission in the nuclei of barred galaxies is predominantly due to nuclear 
\hii\ regions and {\it not} to AGNs.

While it seems well established that bars do in fact help enhance nuclear 
star formation, at least in a statistical sense, their impact on AGN activity 
is far less clear.  Existing observations have yielded somewhat conflicting 
results. This paper aims to elucidate further these issues.  In agreement 
with previous studies, we find that nuclear star formation is enhanced only 
in early-type barred galaxies.  Galaxies hosting AGNs, on the other hand, 
do not appear to be significantly affected by the presence of bars.  We 
argue that AGNs in nearby galaxies might derive most of their power from 
a combination of tidal disruption and stellar mass loss rather than from 
gas accreted from the general interstellar medium of the host galaxy.  
A preliminary discussion of these results was presented by Ho, Filippenko, 
\& Sargent (1996b).  

\section{Data and Analysis}

\subsection{Definition of Sample}

Our analysis draws from a recently completed optical spectroscopic survey 
of the nuclei of a large sample of nearby galaxies (Filippenko \& Sargent
1985; Ho, Filippenko, \& Sargent 1995).  The primary goal of the survey is to 
search for and study low-luminosity AGNs, and it is optimized for the 
detection of very weak emission-line nuclei.  The survey contains 
probably the largest and most complete sample of emission-line nuclei known 
for nearby galaxies (Ho, Filippenko, \& Sargent 1997b) and is especially 
well-suited for addressing the issues at hand.  In brief, high signal-to-noise 
ratio, moderate-resolution (2.5--4 \AA), long-slit spectra were obtained for 
a magnitude-limited ($B_T\,\leq$ 12.5 mag) sample of 486 northern 
(declination $>$ 0\deg) galaxies using the Hale 5-m reflector at Palomar 
Observatory.  Nearly all of the observations were obtained with a 2\asec\ slit, 
and the spectra were extracted using an aperture of 2\asec\ $\times$ 4\asec, 
corresponding to physical dimensions of $\sim$180 pc $\times$ 360 pc for 
the typical distances of the galaxies surveyed (median distance = 17.9 Mpc; 
Ho, Filippenko, \& Sargent 1997a).\footnote{We adopt $H_0$ = 75 \kms\ 
Mpc$^{-1}$ in this series of papers.}  Full details of 
the observations, data reductions, and presentation of the spectra are given 
by Ho \etal (1995), and quantitative measurements of the spectra can be found 
in Ho \etal (1997a).

We wish to examine the effects of bars on the detection frequency of various 
categories of emission-line nuclei.  Adopting the morphological types given in
 the Third Reference Catalogue of Bright Galaxies (RC3; de Vaucouleurs \etal 
1991), our sample contains roughly equal numbers of unbarred (252/486) and 
barred (234/486) galaxies, where unbarred objects are those classified 
as ``A'' in the RC3 and barred ones are those with types ``B'' and ``AB.''  
Barred galaxies comprise 56\% of the disk galaxies (S0--Im) and 59\% of the 
spiral galaxies (S0/a--Sm) in our sample (Ho \etal 1997a).  Our statistics 
agree with those of Sellwood \& Wilkinson (1993), who find that $\sim$60\% of 
field spirals contain bars.  The bar fraction varies with Hubble type.  Spiral 
galaxies of types S0/a to Scd show an approximately constant bar fraction of 
55\%--60\%, while those of type Sd and later have a monotonically 
increasing percentage of barred objects (Fig. 1; see also Table 12 of Ho \etal 
1997a).  Lenticulars (S0) seem to have a somewhat lower incidence of
bars ($\sim$40\%), although the significance of this result is hard to 
judge given the known difficulty of identifying AB systems in this class of 
objects (de Vaucouleurs 1963).  Odewahn (1996) reported very similar trends 
based on a much larger sample of galaxies taken from the RC3.  

We exclude S0 galaxies and consider only spirals in this study.  For reasons 
we do not entirely understand, the barred lenticulars in our sample are 
noticeably less luminous than their unbarred counterparts.  The median 
extinction-corrected absolute blue magnitudes ($M_{B_T}^0$; Ho \etal 1997a) of 
the two groups differ by $\sim$0.9 mag.  We verified that such an offset does 
not exist for each of the spiral types.  The cumulative distributions of 
$M_{B_T}^0$ for the SA0 and SB0+SAB0 samples are significantly different 
according to the Kolmogorov-Smirnov (K-S) test (Press \etal 1986); the 
probability that the two distributions are drawn from the same population 
($P_{\rm KS}$) is 8.5\e{-4}.  Noguchi (1996) surmises that some barred 
lenticulars, specifically those having thin bars, may in fact be late-type 
galaxies depleted of interstellar material.  If true, then this might naturally
explain why the barred sample is somewhat underluminous compared to the 
unbarred sample.  Because the spectral classification of emission-line nuclei 
depends strongly on galaxy Hubble type and luminosity (Ho \etal 1997b), we 
anticipate that the luminosity difference between unbarred and barred 
lenticulars will artificially enhance the AGN fraction in the unbarred 
sample.  Moreover, as has been noted by van den Bergh (1990) and others, the 
S0 class seems to be rather ill-defined and may be especially prone to 
misclassification.  

Our definition of spirals encompasses the Hubble types S0/a through Sm, 
corresponding to T = 0 to 9, where T is the numerical morphological index 
(de Vaucouleurs 1963) as given in the RC3.  The decision to include S0/a and 
Sm (Magellanic spiral) galaxies into the spiral sample is somewhat 
arbitrary and may not be a universally accepted convention.  We chose to do 
so to maximize the sample size, which in total contains 319 objects (132 
unbarred and 187 barred).  We have verified that none of our main conclusions 
are affected by the inclusion of the S0/a and Sm types.

The classifications of the nuclear spectra are taken from Ho \etal (1997a), 
where details of our methodology can be found.  Essentially all the galaxies 
in the spiral sample (97\%) have detectable nuclear emission (Ho \etal 
1997b).  We recognize four classes of emission-line nuclei, defined by the 
relative intensity ratios of several prominent, narrow optical emission lines: 
\hii\ nuclei, Seyfert nuclei, LINERs (low-ionization nuclear emission-line 
regions; Heckman 1980a), and LINER/\hii\ transition objects.  \hii\ nuclei 
spectroscopically resemble \hii\ regions and therefore must be largely 
photoionized by O and B stars.  The other three groups represent variants of 
nuclei most likely powered by nonstellar processes; for the present purposes, 
we will assume that they are all AGNs (see Ho 1996 and Ho, Filippenko, \& 
Sargent 1997d for discussions of the AGN sample).
Note that the majority of the nuclei in question have rather low 
luminosities compared to conventional samples of emission-line nuclei.  Our 
\hii\ nuclei (Ho, Filippenko, \& Sargent 1997c), for instance, emit 
far less line emission than typical ``starburst nuclei,'' and the 
AGNs we analyze have much lower power output than those in most AGN catalogs.

There is an important caveat regarding the utility of our data set for the 
intended application.  Emission arising from either star formation or 
nonstellar activity in the centers of galaxies often extends over a 
considerable area (e.g., Keel 1983a; Pogge 1989).  Barred galaxies of early 
type, in 
particular, favor the formation of nuclear rings which in many cases are 
accompanied by intense star formation (e.g., Telesco, Dressel, \& Woltenscroft
1993).  The circumnuclear rings coincide with the position of the inner 
Lindblad resonance (ILR) and typically extend several hundred parsecs from the 
center.  Thus, two-dimensional imaging or spectra taken through a large 
aperture are required to obtain accurate measurements of the total line 
emission in such systems.  Our spectra, on the other hand, generally probe a 
limited region surrounding the nucleus ($\sim$200 $\times$ 400 pc$^2$) and 
sample only a portion of the emission if it is significantly extended.  
Although the line measurements of individual objects may carry substantial 
uncertainty, we believe that the statistical properties of large numbers 
of objects should be much more reliable, especially when used for comparative 
purposes such as in contrasting barred and unbarred galaxies.

\subsection{Frequency of Nuclear Activity and Presence of a Bar}

We begin by inspecting the dependence of the detection rate of AGNs 
and \hii\ nuclei on the presence of a bar (Table 1 and Fig. 2).   When the 
entire sample of spirals is examined ({\it solid} histograms), it is apparent 
that the frequency of \hii\ nuclei in the barred group is slightly higher than 
that in the unbarred group.  The difference has very low statistical 
significance, although it appears to be real since the effect becomes greater 
when the unbarred group is compared with the strongly-barred types alone 
(i.e., omitting SAB objects).  By contrast, the frequency of AGNs is actually 
somewhat {\it lower} in barred spirals compared to the unbarred spirals.  Once 
again, the trend is more pronounced when we include only the strongly-barred 
types.  The suggested enhancement of the incidence of \hii\ nuclei in 
barred galaxies, and the opposite effect for AGNs, evidently depends on 
morphological type.  While there are not enough objects to examine each Hubble 
type individually, we can form two broad morphological bins: the ``early'' 
group consists of types ranging from S0/a to Sbc, and the ``late'' group spans 
types Sc to Sm.\footnote{A number of studies on barred galaxies (e.g., 
Elmegreen \& Elmegreen 1985, 1989; Combes \& Elmegreen 1993; Huang \etal 1996) 
have found systematic differences in the properties of early-type and late-type 
systems, with the separation occurring roughly at a Hubble type of Sbc.}  In 
the early-type sample ({\it shaded} histograms), no noticeable difference can 
be seen in the percentage of either \hii\ nuclei or AGNs when the barred and 
unbarred galaxies are compared, whereas in the late-type sample ({\it 
unshaded} histograms) the barred galaxies appear to have a higher fraction 
of \hii\ nuclei and a lower fraction of AGNs.  Again, the effect becomes more 
apparent for strongly-barred galaxies.  

Although the apparent differences just mentioned are not large and formally 
carry no statistical significance, we would like to understand their origin 
because they can shed light on possible selection effects present.  Four 
possibilities come to mind.  First, if the barred galaxies on average have a 
larger distance, the fixed slit width of the observations would encompass a 
larger physical region around the nucleus, thereby increasing the probability 
that the ``nuclear'' spectra would be contaminated by emission from 
circumnuclear star-forming regions.  Consequently, for sufficiently strong 
contaminating emission, the detection rate of \hii\ nuclei would increase and 
the detection rate of AGNs would decrease.  However, this explanation cannot 
account for our results, since the distances of the barred and unbarred 
galaxies are very similar, both for the entire sample of spirals and 
individually for the early-type and late-type groups (Table 2).  Second, we 
tested for possible differences in the distribution of absolute magnitudes of 
the barred versus unbarred host galaxies, but none were found.  A third
possibility to note concerns aspect-dependent biases.  We find a highly 
significant difference ($P_{\rm KS}$ = 0.006) in the distributions of 
inclination angles between barred and unbarred spirals (Table 2): unbarred 
objects are more highly inclined (viewed more edge-on) than barred objects by 
6\deg\ in the mean and by 9\deg\ in the median.  The difference persists, 
although at a lower level of statistical significance, when the sample is 
divided into early and late types.  This interesting bias probably derives 
from the original morphological classification of the galaxies; one 
intuitively expects it to be easier to recognize bars in face-on disks than in 
highly inclined ones.  However, a greater fraction of highly inclined disks in 
the unbarred population should result in higher detection rate of \hii \ 
nuclei in this sample, and a correspondingly lower one for AGNs, because of 
the increased probability of intercepting disk or off-nuclear \hii\ regions 
along the line of sight.  Yet, just the opposite is observed, probably because 
the actual differences in inclination angles are relatively minor.  

The fourth and most 
plausible explanation for the variation in detection rates of emission-line 
nuclei between barred and unbarred late-type spirals can be traced to the 
slight differences in the Hubble type distribution in these samples.  We 
already remarked that late-type spirals have a higher bar fraction (Fig. 1).  
The converse is also true: barred spirals have a higher fraction of 
late-type objects compared to unbarred spirals in the same range of 
morphological types.  Table 2 shows that among Sc--Sm spirals, the mean and 
the median T indices of the barred systems are about 0.5 units larger than 
those of the unbarred systems.  This result, while only marginally significant 
according to the K-S test ($P_{\rm KS}$ = 0.095), can probably account for 
the subtle differences in the detection rates of emission-line nuclei.  Since 
\hii\ nuclei occur preferentially in late-type galaxies (Ho \etal 1997b), one 
expects to find a higher fraction of \hii\ nuclei, and a correspondingly lower 
fraction of AGNs, in barred late-type spirals, as observed.

In summary, it appears that the slight enhancement in the frequency of 
\hii\ nuclei among barred late-type spirals, and the opposite trend for 
AGNs, can be largely attributed to small differences in the morphological 
type composition, and {\it not} to the influence of the bar.

\subsection{Strength of Nuclear Activity and Presence of a Bar}

Next, we evaluate the effect of a bar on the {\it strength} of nuclear 
activity.  We use for this purpose the luminosity [L(H\al)] and equivalent 
width [EW(H\al)] of the H\al\ emission line.  The majority of the galaxies 
(75\%) were observed under photometric conditions; Ho \etal (1997a) tabulate 
L(H\al) for these, and they list EW(H\al) for essentially all the 
emission-line objects.\footnote{As discussed by Ho \etal (1997a) and in 
\S\ 2.1, the line measurements refer only to the limited spatial scale sampled 
by the narrow slit of the observations. They are not meant to represent the 
total quantities in the nuclear region.  Broad H\al\ emission is sometimes 
present in AGNs. For a more direct comparison between the AGN and H~II-nuclei 
samples, we do not include this component for the AGNs, but we have checked 
that none of our main conclusions are affected by this choice.}  The 
distribution of L(H\al), corrected for internal extinction, for \hii\ nuclei 
in barred spirals has a tail reaching to much higher luminosities than in 
unbarred spirals (Fig. 3{\it a}; see also Table 3); the cumulative distributions
of the two samples are significantly different according to the K-S test 
($P_{\rm KS}$ = 0.042).  AGNs, on the other hand, do not show such an 
effect (Fig. 3{\it b}); the barred sample does show an enhancement of 
about a factor of two in L(H\al), both in the mean and in the median, but the 
difference between the two distributions has a low statistical significance 
($P_{\rm KS}$ = 0.34).

A surprising trend emerges if we reexamine the data after dividing the sample,
as before, according to Hubble type (Fig. 4; Table 3).  For \hii\ nuclei (Fig. 
4{\it a}), it is apparent that the luminosity enhancement just noted in barred 
galaxies occurs almost entirely in {\it early-type} (S0/a--Sbc) systems.  The 
K-S test indicates that the probability that the L(H\al) distributions of the 
early-type barred and unbarred samples are drawn from the same population 
is $P_{\rm KS}$ = 0.025; for comparison, the same test for the late-type 
objects yields $P_{\rm KS}$ = 0.72.  AGNs (Fig. 4{\it b}) continue not to 
exhibit any pronounced bar-dependent differences.

The equivalent width of the H\al\ emission line potentially serves as a better 
parameter for comparison than L(H\al), since it depends on neither distance 
nor photometric observing conditions (although the previous caveat concerning 
the use of a small aperture still applies).  Removing the latter restriction 
also increases the number of objects available for analysis.  In \hii\ nuclei, 
EW(H\al), defined as the ratio of the flux of H\al\ emission to the continuum 
flux at 6563 \AA, measures the current rate of massive (OB) star formation per 
unit mass, since old stars usually dominate the red continuum.  On the other 
hand, if the nuclei of barred galaxies experience star formation continuously 
over an extended period, the red continuum will be enhanced compared to 
the nuclei of unbarred galaxies, and EW(H\al) may be a less sharp 
star-formation indicator than L(H\al).  In practice, however, this appears not 
to be the case (see below).  In luminous 
AGNs, whose nonstellar optical continuum generally overwhelms the stellar 
background, L(H\al) scales linearly with, and can be considered a measure of,
the luminosity of the nonstellar continuum (e.g., Yee 1980).  However, the red 
continuum in low-luminosity AGNs comes almost entirely from old stars.  If we 
assume for simplicity that photoionization by the nonstellar radiation 
produces all of the H\al\ emission, EW(H\al) in these sources roughly 
indicates the luminosity of the AGN (nonstellar) component relative to the 
normal (stellar) component.  Thus, EW(H\al) can be used as a quantitative 
measure of the relative level of nonstellar activity in our analysis of the 
AGN sample.

As was the case for L(H\al), the distribution of EW(H\al) differs strongly 
between barred and unbarred \hii\ nuclei (Fig. 5{\it a}; $P_{\rm KS}$ = 
0.0092), but not between barred and unbarred AGNs (Fig. 5{\it b}; 
$P_{\rm KS}$ = 0.29).  We further confirm that the enhancement in 
line strength for barred \hii\ nuclei occurs mainly in early-type spirals 
(Fig. 6{\it a}) and that the lack of significant differences between barred 
and unbarred AGNs persists (Fig. 6{\it b}).

While we concluded in \S\ 2.2 that the presence of a bar does not appear to 
increase the likelihood of a galaxy to host either an \hii\ nucleus or an 
AGN, the above analysis suggests that a bar {\it does} enhance the strength or 
intensity of existing nuclear activity, but {\it only} for \hii\ nuclei 
and not for AGNs.  Moreover, the elevation in the rate of current star 
formation appears to favor {\it early-type} spirals.

\section{Interpretation}

\subsection{Bar-Driven Nuclear Star Formation}

Numerical models predict that, even in the absence of other 
non-axisymmetric perturbations such as tidal encounters with neighboring 
companions, a bar can be effective in instigating radial inflow of 
gas toward the center of a galaxy.  In these simulations, 
shocks or cloud-cloud collisions cause gas to pile up along the leading 
edges of the large-scale stellar bar (Roberts, Huntley, \& van Albada 1979; 
Elmegreen 1988).  The resulting asymmetric distribution of the gas with respect 
to that of the stars causes the latter to exert a gravitational torque on 
the former, thereby driving the gas inward.  If the galaxy has an inner 
Lindblad resonance (ILR), the gas will collect near the ILR radius, which 
typically extends several hundred parsecs from the center; in the absence of 
an ILR, the gas can flow closer to the nucleus (e.g., Athanassoula 1992; 
Piner, Stone, \& Teuben 1995).  In either case, gravitational instability in 
the accumulated gas may then lead to a burst of star formation.

Several previous investigations have emphasized the preponderance of 
star-formation activity in the centers of barred galaxies and lend 
observational support to the above theoretical ideas.  Using a relatively 
small sample of nearby galaxies having optical spectra, Heckman (1980b) 
found marginal evidence for enhanced nuclear star formation in barred 
galaxies, at the same time noting that LINERs, galaxies containing compact 
radio cores (which may be regarded as another manifestation of AGNs), and 
Seyfert nuclei apparently behave differently.  That ``peculiar'' or 
``hot-spot'' nuclei containing circumnuclear \hii\ regions (S\'ersic \& 
Pastoriza 1965, 1967) occur almost exclusively in barred galaxies was
noted long ago (S\'ersic \& Pastoriza 1967; S\'ersic 1973; but see the 
criticism by Heckman 1978).  Similarly, Huchra (1977) and Balzano 
(1983) found an excess of barred galaxies among Markarian starbursts.
More recent radio (Hummel \etal 1990) and near-IR (Devereux 1987, 1989) 
studies agree that, while barred galaxies tend to emit more copiously in 
these wavelength bands and most of the emission is confined to the central 
region of the galaxies, the ``activity'' stems mainly from star formation and 
{\it not} from an AGN. 

The results of our analysis add an important dimension to the observational 
picture.  Because our sample is optically selected, it should be much less 
biased toward objects undergoing extreme episodes of star formation than is the 
case for samples chosen by ultraviolet or IR brightness.  Thus, we can better
ascertain the influence of bars in galaxies more representative of the typical 
galaxy population.  The incidence of nuclear star formation in our sample, 
irrespective of the bar type of the host galaxy, is already known to be 
very high (Ho \etal 1997b): over 57\% of all spiral galaxies brighter than 
$B_T$ = 12.5 mag have \hii\ nuclei, with the fraction rising to $\sim$80\% 
in spirals with Hubble types Sc and later.  However, the level of star 
formation is generally quite low; the typical \hii\ nucleus has an H\al\ 
luminosity of only $\sim$2\e{39} \lum\ (Ho \etal 1997b).  This study has 
shown that the presence of a bar in these galaxies on average does enhance the
rate of star formation, as traced by either the luminosity or equivalent 
width of H\al\ emission, but mainly in early-type systems.  On the other hand,
the {\it frequency} of nuclear star formation, as measured by the detection 
rate of \hii\ nuclei, was found to be unchanged by the presence of a bar.  The
minor differences observed in the detection rates were attributable to 
selection effects.  In barred galaxies already containing AGNs, therefore, the 
boosted star formation evidently is insufficiently intense to mask the 
signature of the nonstellar activity.

That the enhancement of nuclear star formation in our sample takes place 
preferentially in 
early-type spirals confirms similar findings reported by Devereux (1987), 
Dressel (1988), and Huang \etal (1996).  As noted by Devereux (1987), the 
dichotomy between the response of the gas to a bar in spirals of early and 
late type probably reflects the influence of the bulge-to-disk ratio on the 
rotation curve and on the relative positions of the primary resonances, as 
the bulge-to-disk ratio is one of the primary parameters that varies along the 
spiral sequence.  The location of the ILR, if present, roughly coincides with 
the turnover radius of the rotation curve (e.g., Combes \& Elmegreen 1993).  
Since galaxies with large bulge-to-disk ratios have steeply-rising rotation 
curves and small turnover radii, an ILR in an early-type disk is expected to 
be located interior to the bar and close to the nucleus; if a late-type disk 
contains an ILR, it will likely be found near the ends of the bar, far from 
the nucleus (Elmegreen \& Elmegreen 1985; Combes \& Elmegreen 1993).  Thus, 
if stars in barred galaxies form preferentially at the ILR, the site where 
the swept-up gas congregates, they are
more likely to be detected as {\it nuclear} star formation in early-type 
systems.  The bulge dominance of a galaxy also influences the length and 
strength of its bar (Elmegreen \& Elmegreen 1985; Combes \& Elmegreen 1993; 
Martin 1995).  Early-type spirals tend to have both longer (relative 
to the galaxy size) and stronger (as parameterized by the bar axial ratio) 
bars than late-type spirals, where the division between ``early'' and ``late'' 
once again occurs roughly at a Hubble type of Sbc.  It is significant 
that recent numerical simulations suggest that the gas accretion rate in 
barred galaxies depends sensitively on the bar strength (Athanassoula 1992; 
Friedli \& Benz 1993): stronger bars can drive more gas to the center.
Indeed, Martin (1995) observes a correlation between the presence of nuclear 
star formation and bar strength.\footnote{Note that Martin's (1995) analysis
partly utilizes the data presented in Arsenault (1989), and therefore may be 
subject to the concerns discussed in the next section.}  As further support 
of this picture, we note that the typical electron density (determined from 
the ratio of \sii\ \lamb 6716 to \sii\ \lamb 6731; see Ho \etal 1997a) in 
the centers of barred early-type galaxies is about a factor of two higher than 
in unbarred objects, whereas in late-type galaxies the difference is 
less conspicuous (Table 3).

It is worth noting that our adopted indicators of star-formation activity 
[L(H\al) and EW(H\al)] span a large range of values both for barred 
{\it and} unbarred galaxies.  This indicates that the presence of a bar is 
neither necessary nor sufficient for nuclear star formation to take place, at 
least not for the levels being considered here.  Undoubtedly other factors 
such as the availability of gas must play a crucial role.  One might argue 
that perhaps galaxies optically classified as unbarred may in fact have bars 
veiled by obscuring dust.  This appears not to always be the case according 
to Pompea \& Rieke (1990), who did not find bars in their near-IR images of 
about 50\% of optically unbarred galaxies having IR characteristics similar to 
those previously shown to have strong nuclear star formation (Hawarden \etal 
1986; Puxley \etal 1988).  The near-IR morphological classification
of Pompea \& Rieke, however, has been criticized recently by Huang \etal 
(1996; see also Hawarden, Huang, \& Gu 1996).  The near-IR survey of paired 
galaxies by Keel, Byrd, \& Klari\'c (1996), on the other hand, similarly finds 
that bars are not ubiquitous, even though one expects bars to be more 
common in interacting systems.  Neglect of the effects of oval 
distortions, whose dynamical influence on the gaseous component of the 
disk is similar to that of a bar (Kormendy 1982), poses another 
concern.  It would be highly desirable to incorporate the effect of ovals 
once sufficiently large, modern photometric surveys of nearby galaxies 
become available.

\subsection{Implications for Fueling of AGNs}

Neither the detection rate nor the emission-line strength of AGNs appears to 
be influenced by the presence of a bar.  This finding applies equally to 
all spirals as well as to early-type and late-type spirals individually.
At first sight, it appears to be at variance with conclusions 
reached by several past studies.  Based on the early observational work of 
Adams (1977) and Simkin \etal (1980), it is sometimes said that AGNs most
frequently occur in barred galaxies (e.g., Shlosman \etal 1989).  It must 
be remembered, however, that both of these early studies were based on 
rather small galaxy samples, and that only a slight preponderance of barred 
host galaxies was suggested.

Using a larger sample of objects, Arsenault (1989) reported 
an overabundance of barred galaxies with inner rings [those classified as 
SAB($r$), SAB($rs$), SB($r$), or SB($rs$)] among \hii\ nuclei {\it and} AGNs.  
Arsenault compared the sample of emission-line nuclei of Keel (1983b) with 
a control sample of galaxies apparently lacking emission-line nuclei selected 
from the survey of Kennicutt \& Kent (1983).  Objects from the control sample 
overlapping with Keel's sample were eliminated, as were those indicated by 
Kennicutt \& Kent as having nuclear H\al\ emission.  This comparison is 
potentially flawed for two reasons.  First, the two samples cannot be properly 
compared because they were observed by different methods.  Keel (1983b) based 
his conclusions on spectra while Kennicutt \& Kent (1983) largely used 
images.  (Kennicutt \& Kent obtained long-slit spectra for only $\sim$30\% of 
the objects in their sample.)  As emphasized by the latter authors, these two 
methods of observation differ in their detection threshold of emission lines, 
with spectroscopy being far more sensitive.  Moreover, Keel (1983b) explicitly 
attempted to correct his spectra for stellar absorption, which further 
enhances the detectability of weak Balmer emission.  Given that nearly all
spiral nuclei exhibit emission lines at some level (Keel 1983b; Ho \etal 
1997b), we seriously doubt that the control sample used by Arsenault (1989) 
truly lacks emission-line nuclei.  The situation is further complicated by the 
fact that those objects apparently lacking nuclear H\al\ emission in the 
Kennicutt \& Kent survey probably have early Hubble types, since one expects 
most of such objects to be AGNs (Ho \etal 1997b) which, on average, have much 
weaker H\al\ emission than \hii\ nuclei (see Fig. 5 and Table 3).  This then 
implies that Arsenault's control sample probably contains a high fraction of 
low-luminosity AGNs.  As a corollary to these complications, a second 
objection to the analysis of Arsenault would be that the two samples being 
compared probably do not have similar distributions of Hubble types, thereby 
introducing unknown selection effects.

In a deep imaging study, Xanthopoulos \& De Robertis (1991) show that 
apparently isolated Seyfert galaxies invariably possess signs of faint 
companions, tidal distortions, or bar-like morphologies.  How to interpret 
these results remains unclear, however, since a sample of non-Seyfert galaxies 
imaged to the same depth is needed for comparison.  

We conclude, therefore, that the apparent lack of impact of bars on 
AGNs does not conflict with existing evidence.  In fact, similar findings, 
albeit based on smaller samples than that available here, have been reported 
by others (Heckman 1980b; Fricke \& Kollatschny 1989; MacKenty 1990).
More recently, McLeod \& Rieke (1995) studied in detail the morphological 
properties of two samples of Seyfert galaxies --- one distilled from 
the CfA redshift survey (Huchra \& Burg 1992) and the other drawn from bright, 
nearby galaxies (Maiolino \& Rieke, unpublished).  They found no excess of 
barred galaxies in either sample.  Moles, M\'arquez, \& P\'erez (1995)
did not find an enhanced bar fraction in a sample of AGNs selected
from the V\'eron-Cetty \& V\'eron  (1991) catalog and for which they 
obtained morphological classifications from the RC3, although they suggested
that there was an excess of inner rings among the subsample of AGNs lacking
bars.  However, it is difficult to assess the significance of the latter result
without a proper control sample.  In their AGN sample, Moles et al. find that 
22\% of the unbarred objects have inner rings.  This does not appear at all 
unusual, for de Vaucouleurs \& Buta (1980) report that the frequency of inner 
rings in unbarred spirals [SA($r$) and SA($rs$)] in the Second Reference 
Catalogue of Bright Galaxies (de Vaucouleurs, de Vaucouleurs, \& Corwin 
1976) is 43\% (for their most reliable sample of angularly large and 
relatively face-on galaxies).  Indeed, it is not obvious why the fraction of 
inner rings in the Moles et al. sample is so {\it low}.  Clearly, updated 
statistics of inner rings from the RC3 are required for a proper comparison, 
and such a study is in progress (Buta 1996).

It has been argued (e.g., Shlosman, Begelman, \& Frank 1990; Heckman 1992) 
that the fraction of barred galaxies may be underestimated in optical 
catalogs, as the effect of dust obscuration is strong and the old stellar 
population (which dominates the mass) does not have a large contrast in 
visible light.  Near-IR imaging surveys have indeed discovered bars in several 
galaxies previously unrecognized as barred at optical wavelengths, but it 
appears that bars are {\it not} universally present in AGNs (McLeod \& Rieke 
1995; see also Mulchaey \& Regan 1997).

The requirements for fueling AGNs are more stringent than those for fueling 
\hii\ nuclei, since angular momentum transport must extend to much smaller 
radii in the former.  Numerical simulations (e.g., Heller \& Shlosman 1994) 
show that a large-scale stellar bar can reduce the angular momentum of the gas 
by only a factor of $\sim$10.  The end product probably leads to 
concentrations of gas, usually of molecular form, observed in the inner 
several hundred parsecs of some barred galaxies (Kenney \etal 1992; Kenney 
1996).  How to further transport the gas to the region of interest for AGNs 
(\lax 1 pc) remains a challenge.  A number of scenarios have been suggested 
(e.g., Lin, Pringle, \& Rees 1988; Shlosman \etal 1989; Hernquist 1989; 
Pfenniger \& Norman 1990; Wada \& Habe 1992, 1995), with the 
``bars-within-bars'' mechanism originally proposed by Shlosman \etal (1989) 
being widely discussed.  When the gaseous disk formed near the nucleus as a 
result of dissipation by the large-scale stellar bar accumulates a 
sufficiently large gas fraction, gravitational instabilities can lead to the 
formation of an inner, secondary gaseous bar.  Analogous to the primary bar, 
the secondary bar drives further inflow, and the whole sequence may be 
repeated on yet smaller scales (Shlosman \etal 1989, 1990).  Although bar-like 
distributions of molecular gas are known in a few nearby galaxies 
(e.g., Young \& Scoville 1991; Kenney 1996), an insufficient number of 
galaxies have been mapped using millimeter-wave interferometers to properly 
assess the statistics of such features.  In a similar spirit, Friedli \& 
Martinet (1993) also invoke the formation of a secondary, nuclear bar as a 
means of transporting matter to small scales, the crucial difference being 
that in their model the secondary bar contains both gas {\it and} stars.  
However, finding nuclear stellar bars is again nontrivial.  Shaw \etal 
(1995) imaged a sample of large barred spirals in the near-IR and concluded 
that only approximately 1/3 of those displaying isophote twists might have a 
genuine nuclear bar.  There are, unfortunately, ambiguities in the 
interpretation of the isophote twists.  A secondary bar could be indicated 
(Friedly \& Martinet 1993; Shaw \etal 1995; Wozniak \etal 1995), but isophote 
twists can also be photometric signatures of an ILR (Shaw \etal 1993; Elmegreen 
\etal 1996) or of a triaxial bulge (Kormendy 1982).

Given that bars aid the radial transport of gas and that AGNs require 
fuel to sustain their activity, how, then, do we understand our result that 
bars apparently do not affect the present level of activity in nearby AGNs?
Let us consider some possible explanations.

(1) A crucial element of all the models which invoke gas dissipation 
to fuel the nucleus is that the gas must constitute a nonnegligible fraction 
of the total dynamical mass in the central region.  For example, in the 
bars-within-bars models, 
the gas fraction amounts to $\sim$10\%--20\% of the local dynamical mass; 
likewise, the self-gravitating nuclear disk envisioned by Lin \etal (1988) 
and Wada \& Habe (1995) requires a comparable amount of gas.  Since AGNs occur 
predominantly in host of early to intermediate types, perhaps the centers of 
these spirals lack sufficient gas for efficient radial transport to take 
place on nuclear scales.  In the few early-type barred spirals which have been 
studied with sufficient angular resolution, many have molecular gas 
distributions peaking near the ILR (Young \& Scoville 1991; Kenney \etal 1992). 
$N$-body simulations (e.g., Combes \& Elmegreen 1993) show that an ILR
naturally develops in early-type barred galaxies, and that, unlike in late-type
barred spirals, it is located interior to the bar.  The large-scale
radial inflow slows down in the vicinity of the ILR, and the material piles up
between the inner and outer ILRs (Athanassoula 1992).  The fate of the gas
accumulated in this region is unclear, as mechanisms for driving the gas either
inward or outward have been proposed (see discussion in Kenney, Carlstrom, \& 
Young 1993).  The apparent lack of increased fueling among the barred AGNs
in our study perhaps favors the latter possibility.  Thus, we speculate that 
the reason a bar does not boost the activity of the active nucleus is that 
the gas never reaches the center, or perhaps does so at a very slow rate.  
The gas content in the central ``cavity'' in barred  early-type galaxies 
may be so low that the instability mechanisms proposed for further inflow 
cannot operate.  A meaningful discussion of this point, as well as 
an observational test of the theoretical models, would require a sensitive, 
high-resolution molecular-line survey of a representative sample of barred 
and unbarred spirals spanning a range of Hubble types.  

(2) Ironically, the accumulation of a central mass concentration by the action 
of a bar (or simply the presence of a supermassive black hole) and the 
concomitant development of a strong and extended ILR can also lead to the {\it 
destruction} of the bar (Hasan \& Norman 1990; Pfenniger \& Norman 1990; 
Friedli \& Benz 1993).  Such a regulatory mechanism was proposed 
by Friedli (1994) to account for the scarcity of luminous AGNs in the 
current epoch.  Although this phenomenon may explain why the frequency of 
barred and unbarred AGNs is the same, it does not resolve the quandary that 
barred and unbarred AGNs currently show activity at the same level.  Since the 
survivability of the bar is thought to depend sensitively on the mass of the 
central concentration relative to the mass of the stellar disk (Friedli 1994), 
this scenario may be testable by appropriate kinematic observations.

(3) One might appeal to a duty cycle argument.  Suppose that bar-induced 
inflow is not quasi-continuous, but rather episodic in nature, with the 
duration between episodes of accretion, or the latency period, lasting 
on the order of the lifetime of the bar.  A similar idea was advanced by 
Byrd, Sundelius, \& Valtonen (1987) to explain the paucity of Seyfert nuclei
among galaxies in multiple systems.  In the present context, the latency 
period might be related to the timescale for accumulating a critical amount 
of gas to initiate the instabilities required for the various mechanisms of 
radial transport to the nucleus, in conjunction with accretion time 
itself.  In the absence of conditions which lead to 
bar destruction, bars appear to be long-lived structures (Sellwood \& 
Wilkinson 1993).  If the duty-cycle argument holds, it implies that fueling 
of nearby AGNs on nuclear dimensions must be very slow (e.g., because gas is 
deficient on the relevant physical scale) or that the proposed mechanisms of 
radial inflow are more inefficient than assumed.

The considerations discussed above can be bypassed altogether if one dispenses
with the notion that gas from the kiloparsec-scale region is the principal 
source of fuel in AGNs.  The idea that fuel may be derived from tidal 
disruption of {\it stars} by the central source (presumably a supermassive 
black hole) was proposed by Hill (1975) and has subsequently been developed 
extensively (see Rees 1988, Shlosman \etal 1990, and Roos 1992 for reviews).  
The viability of this mechanism has been criticized on several grounds 
(Shlosman \etal 1990), principally that the fueling rate required to sustain 
the prodigious output of luminous QSOs may be too large and that star clusters 
more compact than have yet been detected are required.  Shlosman \etal (1990) 
further remark that if such compact clusters existed, one would still need to 
cope with the problem of angular momentum transport implicit in their 
formation.  On the other hand, these objections do not apply to AGNs of 
{\it low luminosity}, such as the ones populating nearby galaxies and being 
considered here.  For an efficiency of conversion between matter and energy of 
$\epsilon$ = 0.1, the mass accretion rate required to sustain a luminosity 
L is $\dot M$ = $(\epsilon\,c^2)^{-1} L\,=\,0.15\,(\epsilon/0.1)\,(L/10^{45}\,
{\rm ergs\, s^{-1}})$ \solmass\ \peryr.  While QSOs typically consume 
$\dot M\,\approx$ 10--100 \solmass\ \peryr, a Seyfert 2 nucleus such as 
that of NGC~1068 (Pier \etal 1994) only requires $\sim$0.2 \solmass\ \peryr, 
the LINER/Seyfert nucleus in M81 just $\sim$5\e{-5} \solmass\ \peryr\ (Ho, 
Filippenko, \& Sargent 1996a), and, in the extreme case of the Seyfert 1 
nucleus in NGC~4395 (Filippenko \& Sargent 1989), a factor of 10 lower 
than the nucleus of M81.  According to Eracleous, Livio, \& Binette (1995), 
one could expect a stellar disruption once every 100--200 years for a black 
hole mass of 10$^6$--10$^7$ \solmass; the tidal debris forms an accretion disk 
capable of sustaining the ionizing radiation of low-luminosity sources for 
several decades.  (These authors considered LINERs, but the exact type of 
AGN is irrelevant, as long as they have low luminosities.)  Their estimates are 
based on the observed central stellar densities of nearby galaxies and 
do not depend on the existence of the hypothetical clusters needed for more 
luminous sources.  

In most nearby AGNs, it can be shown that mass loss from the evolution of 
normal stars in the vicinity of the nucleus can also provide an appreciable 
amount of fuel.  By only considering stars in the central few parsecs around
the nucleus, angular momentum transfer at large scales can be neglected.
Modeling the amount of gas retained by a galaxy during the course of its 
evolution depends on a number of poorly understood processes (Padovani 
\& Matteucci 1993).  Using a self-consistent model which successfully 
reproduces many of the observed properties of elliptical galaxies, Padovani 
\& Matteucci calculated the time dependence of the stellar mass loss rate on 
the total galaxy luminosity.  For a Salpeter initial mass function with 
lower and upper mass cutoffs of 0.1 and 100 \solmass, respectively, a 15 Gyr 
system is expected to lose mass at a rate of $\dot M_*\,\approx$ 3\e{-11} 
(L/\solum) \solmass\ \peryr, where L is measured in the $V$ band; thus, 
$\dot M_*$ scales linearly with the luminosity.  {\it Hubble Space Telescope 
(HST)} images with spatial resolution limits of a few parsecs show that the
central regions of many galaxies typically have luminosity densities of 
$\sim$10$^3$--10$^4$ \solum\ pc$^{-3}$ ($\sim$$V$ band; Lauer \etal 1995).  
Assuming that the Padovani \& Matteucci calculations can be applied to the 
nuclei of early-type spirals (whose stellar population is generally old, and, 
as a rough approximation, resembles that of elliptical galaxies), it seems
that mass loss from evolved stars in the central $\sim$5 pc can provide most 
of the fuel for AGNs like M81.  A handful of the galaxies imaged by Lauer 
\etal in addition were found to possess nuclear star clusters with 
luminosities of $\sim$10$^6$--10$^7$ \solum\ (within the central few parsecs); 
if such clusters exist around low-luminosity AGNs, they can also furnish a 
substantial fraction of the fuel.  [In these estimates, we make the simple 
assumption that {\it all} of the gas released through mass loss within the 
central few parsecs gets accreted.  In low-luminosity AGNs such as M81 
these dimensions roughly correspond to the inner portion of the 
narrow-line region.  In reality, the fraction of the gas actually captured and 
ultimately accreted is unclear.]  Whether such high stellar densities are 
commonplace in galactic nuclei and whether galaxies hosting AGNs 
preferentially have higher central stellar densities remain to be seen.  We 
are currently pursuing these issues by systematically studying with the 
{\it HST} the morphologies of the central regions of a large sample of 
nearby galaxies selected from our ground-based survey.

As a first approximation, it appears that the normal stellar component of 
galactic nuclei alone, either through stellar capture by a supermassive black 
hole or in combination with stellar mass loss, can supply enough fuel for 
low-luminosity AGNs.  Nevertheless, even in this extreme scenario which 
neglects the contribution from dissipation of the interstellar component, it 
remains surprising that bars have no apparent effect on the fueling rate.  
In the presence of a non-axisymmetric gravitational perturbation, the 
disruption rate of stars should still be significantly boosted.  While 
in a spherically symmetric potential the replenishment of the stellar-orbit 
loss cone is limited by the two-body relaxation timescale, Norman \& Silk 
(1983) demonstrate that the loss cone can be refueled much more efficiently 
in a triaxial gravitational potential.  The resulting stellar capture rate 
may be as much as a factor of ten higher.

\section{Conclusions}

Using a large sample of emission-line nuclei derived from a 
spectroscopic survey of nearby galaxies, we investigate the effect of a 
bar on nuclear star formation and nonstellar activity.  Specifically, 
we focus on the question of whether the frequency and strength of these two 
types of activity are enhanced in barred spirals compared to unbarred spirals.
Numerical simulations indicate that the non-axisymmetric 
gravitational perturbation of a bar in a disk galaxy serves as an effective 
agent to remove angular momentum of the gas on large scales.  Gas flows 
toward the central (\lax 1 kpc) region of the galaxy, where rapid star 
formation may be initiated.  Further inflow to physical scales of 
relevance for active nuclei may be possible according to different 
theoretical suggestions.

The incidence of star-forming (\hii) nuclei is marginally enhanced in barred 
(relative to unbarred) late-type spirals (Sc--Sm); this is accompanied by a 
slight decrease in the detection rate of AGNs in the same range of Hubble 
types.  These small differences, however, which carry no statistical 
significance, can be attributed to the tendency for the barred sample 
to be somewhat skewed toward later Hubble types.  The presence of a bar has no 
noticeable effect on the likelihood of galaxies to host nuclear activity.  
On the other hand, bars do seem to boost the rate of star formation as 
measured by either the luminosity or the equivalent width of H\al\ emission, 
although the effect is observed primarily in early-type spirals (S0/a--Sbc).
The enhancement of star formation in the centers of early-type barred spirals,
previously noted in several studies, can be explained most easily in 
terms of structural differences of the bars in early and late systems.  An 
early-type system, with a large bulge-to-disk ratio, is expected to have 
an inner Lindblad resonance interior to the bar and close to the nucleus; in 
a late-type system, if the resonance exists at all, it is located near the 
ends of the bar and far from the nucleus.  Gas in barred galaxies tends to 
accumulate near the resonance, and, under suitable conditions, stars may 
form.  Thus, nuclear star formation should be preferentially seen in 
early-type systems.  Furthermore, the relatively larger strength and 
dimension of bars in early-type systems can lead to greater gas inflow rates, 
further increasing the probability of star formation in the galaxy centers.
Nuclear star formation spans a wide range of intensity in both barred and 
unbarred galaxies.  Although the presence of a bar on average has a positive 
impact on star formation in the centers of early-type spirals, it should be 
remembered that it is neither a necessary nor a sufficient condition for 
star formation to occur.

Galaxies hosting active nuclei behave quite differently.  Bars have no 
obvious impact on the strength of the AGN, reaffirming 
similar conclusions reached by other studies based on smaller samples.
If one postulates that the interstellar medium of the host galaxy provides 
the bulk of the fuel to AGNs, these results imply at least one of the 
following: (1) The amount of gas present in the central regions of early-type 
galaxies (the dominant hosts of AGNs) is insufficient to sustain the various 
instability mechanisms proposed for radial transport of material to nuclear 
scales; this may be related to the preponderance of inner Lindblad resonances 
in early-type galaxies, as the resonances can obstruct further inflow of gas 
to the nucleus.  (2) The presence of a supermassive black hole in the centers 
of AGN hosts leads to the destruction of the bar.  (3) Bar-induced accretion 
is episodic, with the length of the duty cycle being comparable to the 
lifetime of the bar.  Some of these hypotheses can be tested with future 
observations, but none is entirely satisfactory.

These ambiguities can be largely circumvented if nearby AGNs derive their 
primary fuel supply not from accretion of the large-scale interstellar medium, 
but rather from the stars immediately surrounding the galaxy nuclei.  Although 
fuel from a purely stellar origin faces difficulties in accounting for 
luminous QSOs, we argue that it might adequately sustain the luminosities of 
less powerful sources such as many nearby Seyfert and LINER nuclei.  Tidal 
disruption of stars by a supermassive black hole can feed the central 
engine, as can mass loss from evolved stars in the vicinity of the nucleus.
It remains to be understood, however, why the non-axisymmetric perturbation 
of a bar does not enhance the rate of stellar disruptions.

\acknowledgments
The research of L.~C.~H. is currently funded by a postdoctoral fellowship
from the Harvard-Smithsonian Center for Astrophysics.  Financial support for
this work was provided by NSF grants AST-8957063 and AST-9221365, as well as
by NASA grants AR-5291-93A and AR-5792-94A from the Space Telescope Science
Institute (operated by AURA, Inc., under NASA contract NAS5-26555).
L.~C.~H. thanks Nick Devereux, Masafumi Noguchi, Ron Buta, Aaron Barth, and
Sydney van~den~Bergh for discussions on barred galaxies.  We are grateful 
to Chris McKee and Hy Spinrad for critically reading an earlier draft of the 
manuscript, and to Bill Keel, the referee, for helpful comments.

\clearpage

\centerline{\bf{References}}
\medskip

\refindent
Adams, T.~F. 1977, \apjs, 33, 19

\refindent
Arsenault, R. 1989, \aa, 217, 66

\refindent
Athanassoula, E. 1992, \mnras, 259, 345

\refindent
Balzano, V.~A. 1983, \apj, 268, 602

\refindent
Barnes, J.~E., \& Hernquist, L.~E. 1992, \annrev, 30, 705

\refindent
Buta, R.~J. 1996, private communication

\refindent
Byrd, G., Sundelius, B., \& Valtonen, M. 1987, \aa, 171, 16

\refindent
Combes, F., \&  Elmegreen, B.~G. 1993, \aa, 271, 391

\refindent
de Jong, T., \etal 1984, \apj, 278, L67

\refindent
de Vaucouleurs, G. 1963, \apjs, 8, 31

\refindent
de Vaucouleurs, G., \& Buta, R.~J. 1980, \apjs, 44, 451

\refindent
de Vaucouleurs, G., de Vaucouleurs, A., \& Corwin, H.~G., Jr. 1976, Second
Reference Catalogue of Bright Galaxies (Austin: Univ. of Texas Press) 

\refindent
de Vaucouleurs, G., de Vaucouleurs, A., Corwin, H.~G., Jr., Buta, R.~J.,
Paturel, G., \& Fouqu\'e, R. 1991, Third Reference Catalogue of Bright
Galaxies (New York: Springer) (RC3)

\refindent
Devereux, N.~A. 1987, \apj, 323, 91

\refindent 
Devereux, N.~A. 1989, \apj, 346, 126

\refindent
Dressel, L.~L. 1988, \apj, 329, L69

\refindent
Elmegreen, B.~G. 1988, \apj, 326, 616


\refindent
Elmegreen, B.~G., \& Elmegreen, D.~M. 1985, \apj, 288, 438

\refindent
Elmegreen, B.~G., \& Elmegreen, D.~M. 1989, \apj, 342, 677

\refindent
Elmegreen, D.~M., Elmegreen, B.~G., Chromey, F.~R., Hasselbacher, D.~A., \&
Bissell, B.~A. 1996, \aj, 111, 1880

\refindent
Eracleous, M., Livio, M., \& Binette, L. 1995, \apj, 445, L1

\refindent
Filippenko, A.~V., \& Sargent, W.~L.~W. 1985, \apjs, 57, 503

\refindent
Filippenko, A.~V., \& Sargent, W.~L.~W. 1989, \apj, 342, L11

\refindent
Fricke, K.~J., \& Kollatschny, W. 1989, in Active Galactic Nuclei, ed.
D.~E. Osterbrock \& J.~S. Miller (Dordrecht: Kluwer), 425

\refindent
Friedli, D. 1994, in Mass Transfer Induced Activity in Galaxies, ed.  I.
Shlosman (Cambridge: Cambridge Univ. Press), 268

\refindent
Friedli, D., \& Benz, W. 1993, \aa, 268, 65

\refindent
Friedli, D., \& Martinet, L. 1993, \aa, 227, 27

\refindent
Hasan, H., \& Norman, C.~A. 1990, \apj, 361, 69

\refindent
Hawarden, T.~G., Huang, J.~H., \& Gu, Q.~S. 1996, in IAU Colloq. 157, Barred
Galaxies, ed. R. Buta, B.~G. Elmegreen, \& D.~A. Crocker (San Francisco:
ASP), 54

\refindent
Hawarden, T.~G., Mountain, C.~M., Leggett, S.~K., \& Puxley, P.~J. 1986,
\mnras, 221, 41P

\refindent
Heckman, T.~M. 1978, \pasp, 90, 241

\refindent
Heckman, T.~M. 1980a, \aa, 87, 152

\refindent
Heckman, T.~M. 1980b, \aa, 88, 365

\refindent
Heckman, T.~M. 1992, in Testing the AGN Paradigm, ed. S. Holt, S. Neff, \& M.
Urry (New York: AIP), 595

\refindent
Heckman, T.~M. 1994, in Mass Transfer Induced Activity in Galaxies, ed.
I. Shlosman (Cambridge: Cambridge Univ. Press), 234

\refindent
Heller, C.~H., \& Shlosman, I. 1994, \apj, 424, 84

\refindent
Hernquist, L. 1989, \nat, 340, 687

\refindent
Hill, J.~G. 1975, \nat, 254, 295

\refindent
\refindent
Ho, L.~C. 1996, in The Physics of LINERs in View of Recent Observations, ed.
M. Eracleous, et al. (San Francisco: ASP), 103
 

\refindent
Ho, L.~C., Filippenko, A.~V., \& Sargent, W.~L.~W. 1995, \apjs, 98, 477
 
\refindent
Ho, L.~C., Filippenko, A.~V., \& Sargent, W.~L.~W. 1996a, \apj, 462, 183
 
\refindent
Ho, L.~C., Filippenko, A.~V., \& Sargent, W.~L.~W. 1996b, in IAU Colloq. 157,
Barred Galaxies, ed. R. Buta, B.~G. Elmegreen, \& D.~A. Crocker (San
Francisco: ASP), 188

\refindent
Ho, L.~C., Filippenko, A.~V., \& Sargent, W.~L.~W. 1997a, \apjs, in press

\refindent
Ho, L.~C., Filippenko, A.~V., \& Sargent, W.~L.~W. 1997b, \apj, in press
 
\refindent
Ho, L.~C., Filippenko, A.~V., \& Sargent, W.~L.~W. 1997c, \apj, in press

\refindent
Ho, L.~C., Filippenko, A.~V., \& Sargent, W.~L.~W. 1997d, in preparation

\refindent
Huang, J.~H., Gu, Q.~S., Su, H.~J., Hawarden, T.~G., Liao, X.~H., \& Wu, G.~X.
1996, \aa, 313, 13

\refindent
Huchra, J.~P. 1977, \apjs, 35, 171

\refindent
Huchra, J.~P., \& Burg, R. 1992, \apj, 393, 90

\refindent
Hummel, E., van der Hulst, J.~M., Kennicutt, R.~C., Jr., Keel, W.~C. 1990,
\aa, 236, 333

\refindent
Keel, W.~C. 1983a, \apj, 268, 632

\refindent
Keel, W.~C. 1983b, \apj, 269, 466

\refindent
Keel, W.~C., Byrd, G.~G., \& Klari\'c, M. 1996, in IAU Colloq. 157, Barred 
Galaxies, ed.\ R.  Buta, B. G. Elmegreen, \& D. A. Crocker (San Francisco: 
ASP), 360

\refindent
Kenney, J.~D. 1996, in IAU Colloq. 157, Barred Galaxies, ed. R. Buta, B.~G.
Elmegreen, \& D.~A. Crocker (San Francisco: ASP), 150

\refindent
Kenney, J.~D., Carlstrom, J.~E., \& Young, J.~S. 1993, \apj, 418, 687

\refindent
Kenney, J.~D.~P., Wilson, C.~D., Scoville, N.~Z., Devereux, N.~A., \& Young,
J.~S. 1992, \apj, 395, L79

\refindent
Kennicutt, R.~C., \& Kent, S.~M. 1983, \aj, 88, 1094

\refindent
Kormendy, J. 1982, in Morphology and Dynamics of Galaxies, ed. L. Martinet \& 
M. Mayor (Geneva: Geneva Observatory), 115

\refindent
Lauer, T.~R., \etal 1995, \aj, 110, 2622

\refindent
Lin, D.~C., Pringle, J.~E., \& Rees, M.~J. 1988, \apj, 328, 103

\refindent 
MacKenty, J.~W. 1990, \apjs, 72, 231

\refindent
Martin, P. 1995, \aj, 109, 2428

\refindent
McLeod, K.~K., \& Rieke, G.~H. 1995, \apj, 441, 96

\refindent
Moles, M., M\'arquez, I., \& P\'erez, E. 1995, \apj, 438, 604

\refindent
Mulchaey, J.~S., \& Regan, M.~W. 1997, \apj, in press

\refindent
Noguchi, M. 1996, \apj, 469, 605

\refindent
Norman, C., \& Silk, J. 1983, \apj, 266, 502

\refindent
Odewahn, S.~C. 1996, in IAU Colloq. 157, Barred Galaxies, ed. R. Buta, B.~G. 
Elmegreen, \& D.~A. Crocker (San Francisco: ASP), 30

\refindent
Padovani, P., \& Matteucci, F. 1993, \apj, 416, 26

\refindent
Pfenniger, D., \& Norman, C. 1990, \apj, 363, 391

\refindent
Phinney, E.~S. 1994, in Mass Transfer Induced Activity in Galaxies, ed.
I. Shlosman (Cambridge: Cambridge Univ. Press), 1

\refindent
Pier, E.~A., Antonucci, R., Hurt, T., Kriss, G., \& Krolik, J. 1994, \apj,
  428, 124

\refindent
Piner, B.~G., Stone, J.~M., \& Teuben, P.~J. 1995, \apj, 449, 508

\refindent
Pogge, R.~W. 1989, \apjs, 71, 433

\refindent
Pompea, S.~M., \& Rieke, G.~H. 1990, \apj, 356, 416

\refindent
Press, W.~H., Flannery, B.~P., Teukolsky, S.~A., \& Vetterling, W.~T. 1986,
Numerical Recipes (Cambridge: Cambridge Univ. Press)

\refindent
Puxley, P.~J., Hawarden, T.~G., \& Mountain, C.~M. 1988, \mnras, 231, 465

\refindent
Rees, M.~J. 1988, \nat, 333, 523

\refindent
Roberts, W.~W., Jr., Huntley, J.~M., \& van Albada, G.~D. 1979, \apj, 233, 67

\refindent
Roos, N. 1992, \apj, 385, 108

\refindent
Sellwood, J.~A., \& Wilkinson, A. 1993, Rep. Prog. Phys., 56, 173

\refindent
S\'ersic, J.~L. 1973, \pasp, 85, 103

\refindent
S\'ersic, J.~L., \& Pastoriza, M. 1965, \pasp, 77, 287

\refindent
S\'ersic, J.~L., \& Pastoriza, M. 1967, \pasp, 79, 152

\refindent
Shaw, M., Combes, F., Axon, D., \& Wright, G.~S. 1993, \aa, 273, 31

\refindent
Shaw, M., Axon, D., Probst, R., \& Gatley, I. 1995, \mnras, 274, 369

\refindent
Shlosman, I., ed., Mass Transfer Induced Activity in Galaxies (Cambridge: 
Cambridge Univ. Press)

\refindent
Shlosman, I., Begelman, M.~C., Frank, J. 1990, \nat, 345, 679

\refindent
Shlosman, I., Frank, J., \& Begelman, M.~C. 1989, \nat, 338, 45

\refindent
Simkin, S.~M., Su, H.~J., \& Schwarz, M.~P. 1980, \apj, 237, 404

\refindent
Sulentic, J.~W., Keel, W.~C., \& Telesco, C.~M., eds., Paired and Interacting 
Galaxies (NASA CP-3098)

\refindent
Telesco, C.~M., Dressel, L.~L., \& Woltenscroft, R.~D. 1993, \apj, 414, 120

\refindent
van den Bergh, S. 1990, \apj, 346, 57

\refindent
V\'eron-Cetty, M.-P., \& V\'eron, P. 1991, A Catalog of Quasars and Active
Nuclei (ESO Scientific Rep. 10)
 
\refindent
Wada, K., \& Habe, A. 1992, \mnras, 258, 82

\refindent
Wada, K., \& Habe, A. 1995, \mnras, 277, 433

\refindent
Xanthopoulos, E., \& De Robertis, M.~M. 1991, \aj, 102, 1980

\refindent
Yee, H.~K.~C. 1980, \apj, 241, 894

\refindent
Young, J.~S., \& Scoville, N.~Z. 1991, \annrev, 29, 581

\refindent
Wozniak, H., Friedli, D., Martinet, L., Martin, P., \& Bratschi, P. 1995,
\aas, 111, 115

\clearpage
\begin{figure}
\plotone{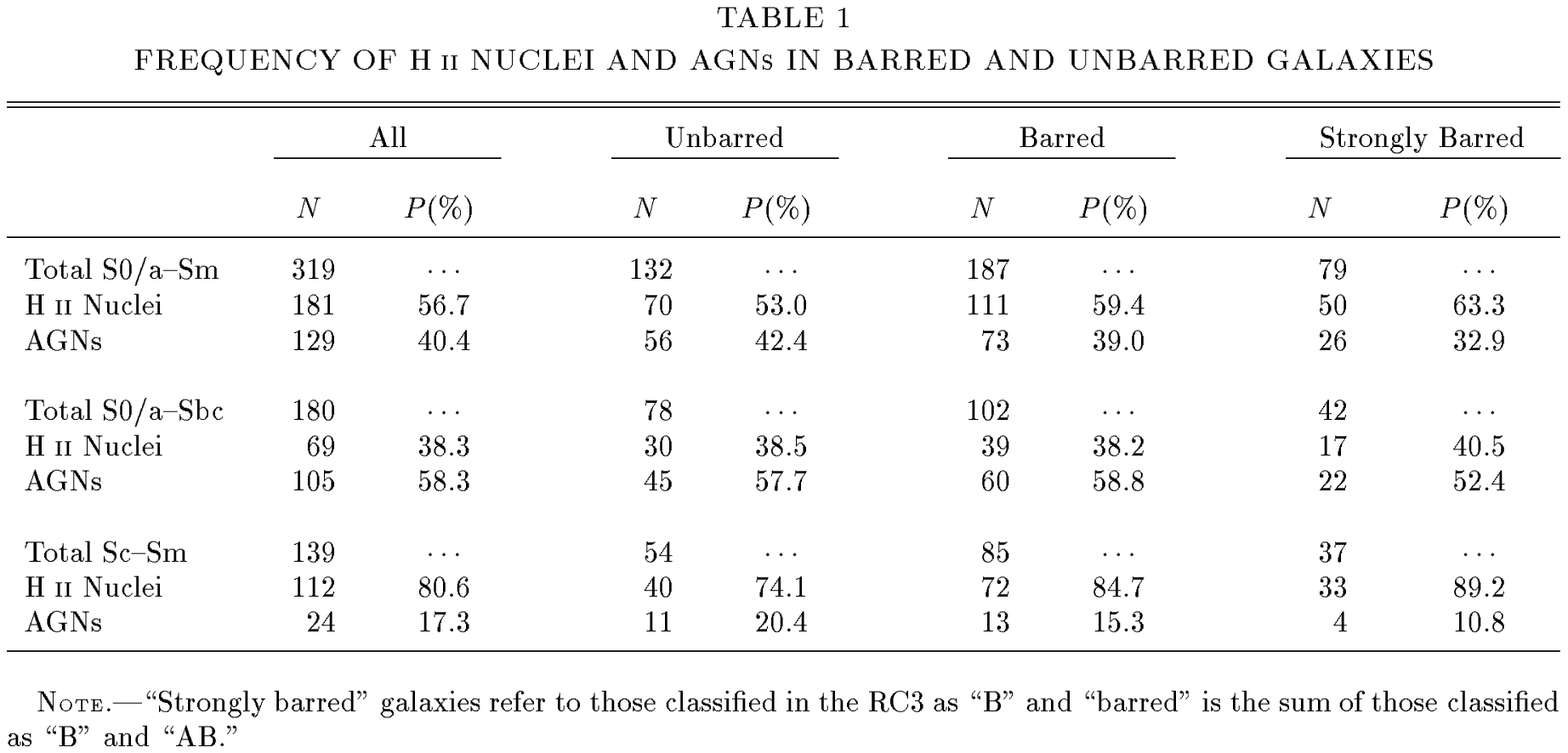}
\end{figure}

\clearpage
\begin{figure}
\plotone{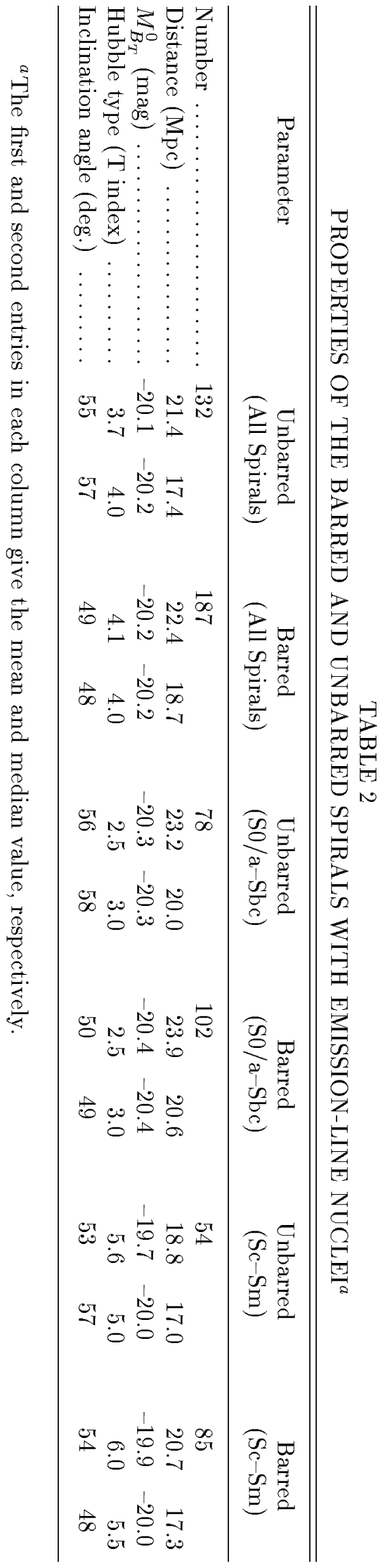}
\end{figure}

\clearpage
\begin{figure}
\plotone{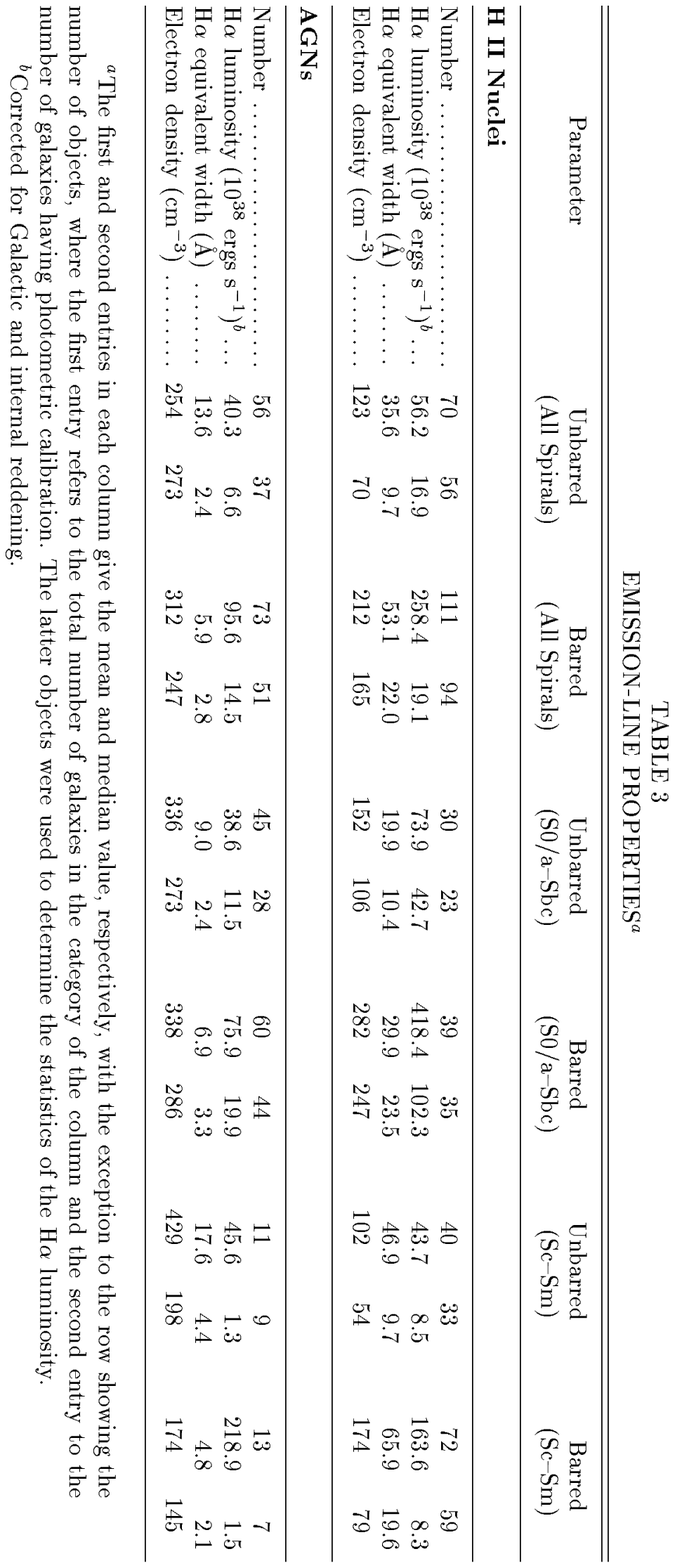}
\end{figure}

\clearpage

\centerline{\bf{Figure Captions}}
\medskip

Fig. 1. ---
The percentage of barred galaxies as a function of Hubble type for 417 
lenticular (S0) and spiral (S0/a--Sm) galaxies. The {\it dotted} line shows 
the weakly-barred systems (SAB), the {\it dashed} line the strongly-barred 
systems (SB), and the {\it solid} line both types combined (SAB+SB).  
The error bars simply reflect counting statistics and are given by 
$[f(1-f)/N]^{1/2}$, where $f$ is the fraction of objects belonging to a 
particular bin and $N$ is the total number of objects in that bin.  The three 
groups are slightly shifted along the abscissa for clarity.  The 
bottom scale gives the Hubble types, and the top scale the corresponding 
numerical T indices.  The types were slightly binned as follows:  
``S0'' = S0, ``Sa'' = S0/a--Sab, ``Sb'' = Sb--Sbc, ``Sc'' = Sc--Scd, 
``Sd'' = Sd--Sdm, and ``Sm'' = Sm.

Fig. 2. ---
Dependence of the detection rates of \hii\ nuclei and AGNs on the morphology 
of spiral galaxies (Table 1).  The group labeled ``strongly barred'' includes 
only galaxies classified as ``SB,'' while the group labeled ``barred'' 
includes both ``SB'' and ``SAB'' classifications.  The {\it solid} 
histograms include all spirals (S0/a--Sm), the {\it shaded} histograms only 
early-type spirals (S0/a--Sbc), and the {\it unshaded} histograms only 
late-type (Sc--Sm) spirals.

Fig. 3. ---
Distribution of extinction-corrected H\al\ luminosities (Ho \etal 1997a) for
({\it a}) \hii\ nuclei and ({\it b}) AGNs.  The top and bottom panels
show barred and unbarred galaxies, respectively. The bins are separated by 
0.25 in logarithmic units.

Fig. 4. ---
Distribution of extinction-corrected H\al\ luminosities for ({\it a}) \hii\ 
nuclei and ({\it b}) AGNs.  The two top panels in each case show barred and
unbarred galaxies for early-type spirals (S0/a--Sbc), and the two bottom 
panels show late-type spirals (Sc--Sm). The bins are separated by 0.25 in 
logarithmic units.

Fig. 5. ---
Distribution of H\al\ equivalent widths (Ho \etal 1997a) for ({\it a}) \hii\ 
nuclei and ({\it b}) AGNs.  The top and bottom panels show barred and unbarred 
galaxies, respectively.  The bins are separated by 2 \AA, and the last bin 
contains all objects with EW(H\al) $>$ 30 \AA.

Fig. 6. ---
Distribution of H\al\ equivalent widths for ({\it a}) \hii\ nuclei and 
({\it b}) AGNs.  The two top panels in each case show barred and unbarred 
galaxies for early-type (S0/a--Sbc) spirals, and the two bottom panels show 
late-type (Sc--Sm) spirals.  The bins are separated by 2 \AA, and the last bin 
contains all objects with EW(H\al) $>$ 30 \AA.

\clearpage
\begin{figure}
\figurenum{1}
\plotone{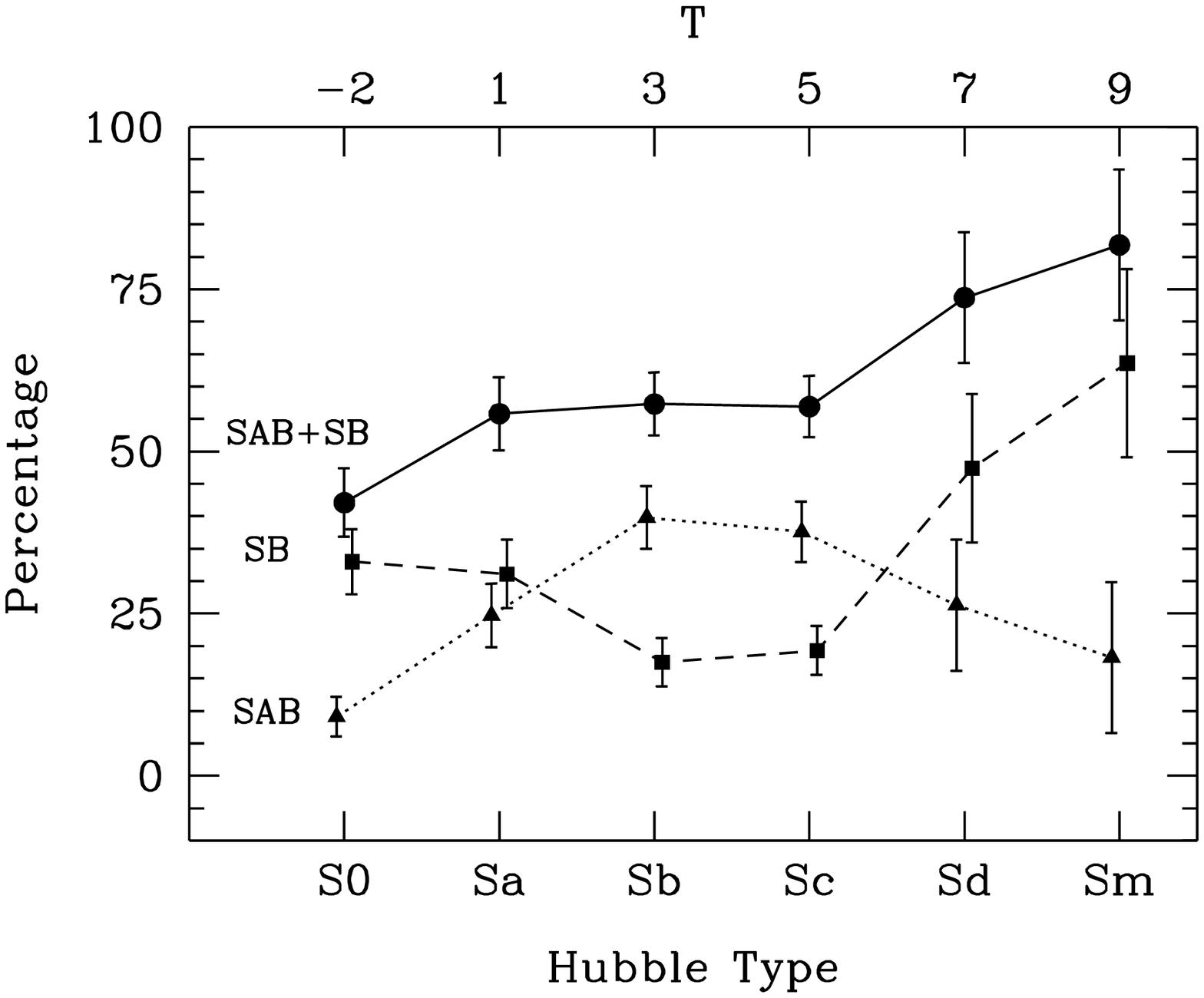}
\caption{}
\end{figure}

\clearpage
\begin{figure}
\figurenum{2}
\plotone{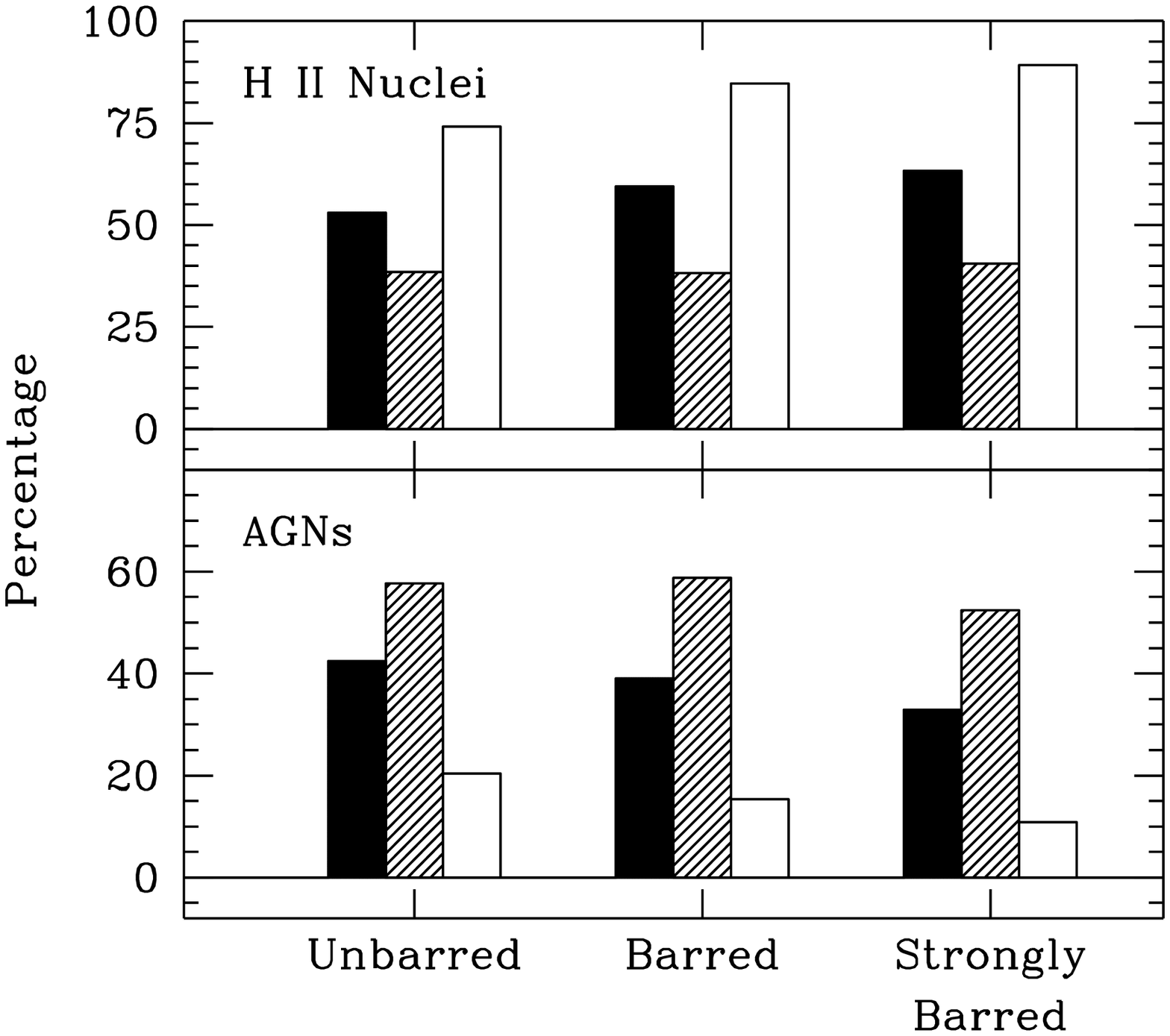}
\caption{}
\end{figure}

\clearpage
\begin{figure}
\figurenum{3}
\plotone{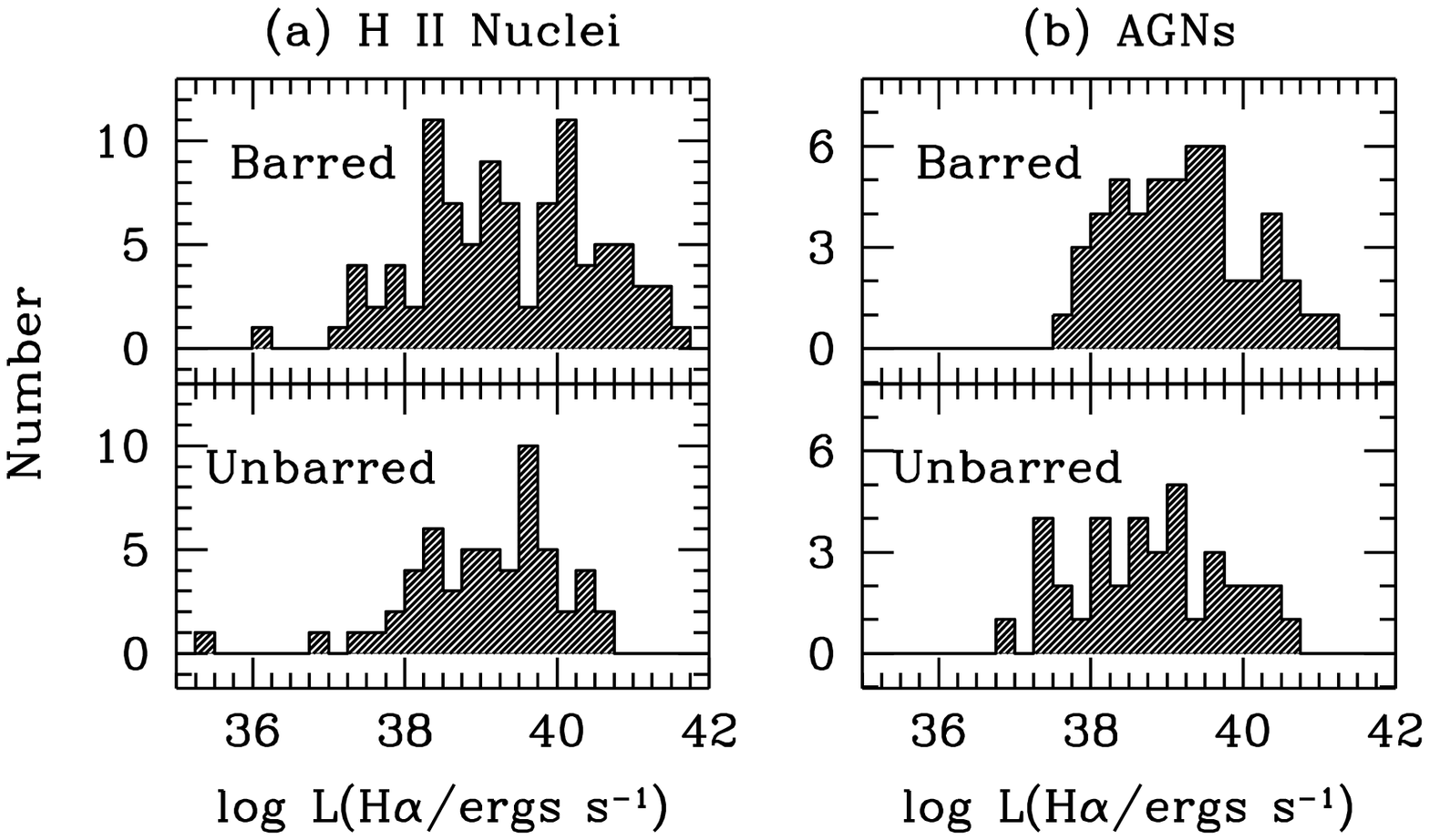}
\caption{}
\end{figure}

\clearpage
\begin{figure}
\figurenum{4}
\plotone{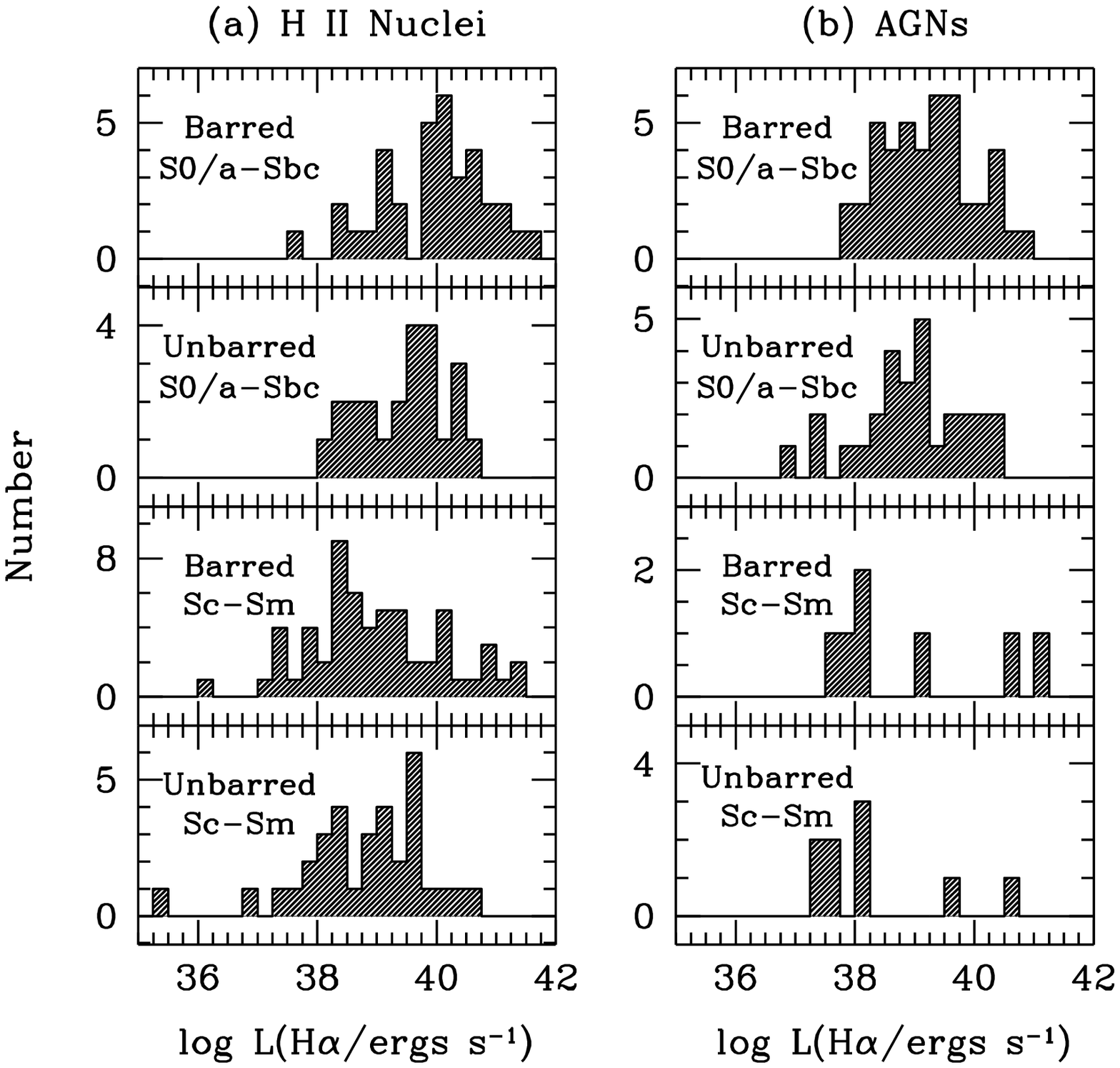}
\caption{}
\end{figure}

\clearpage
\begin{figure}
\figurenum{5}
\plotone{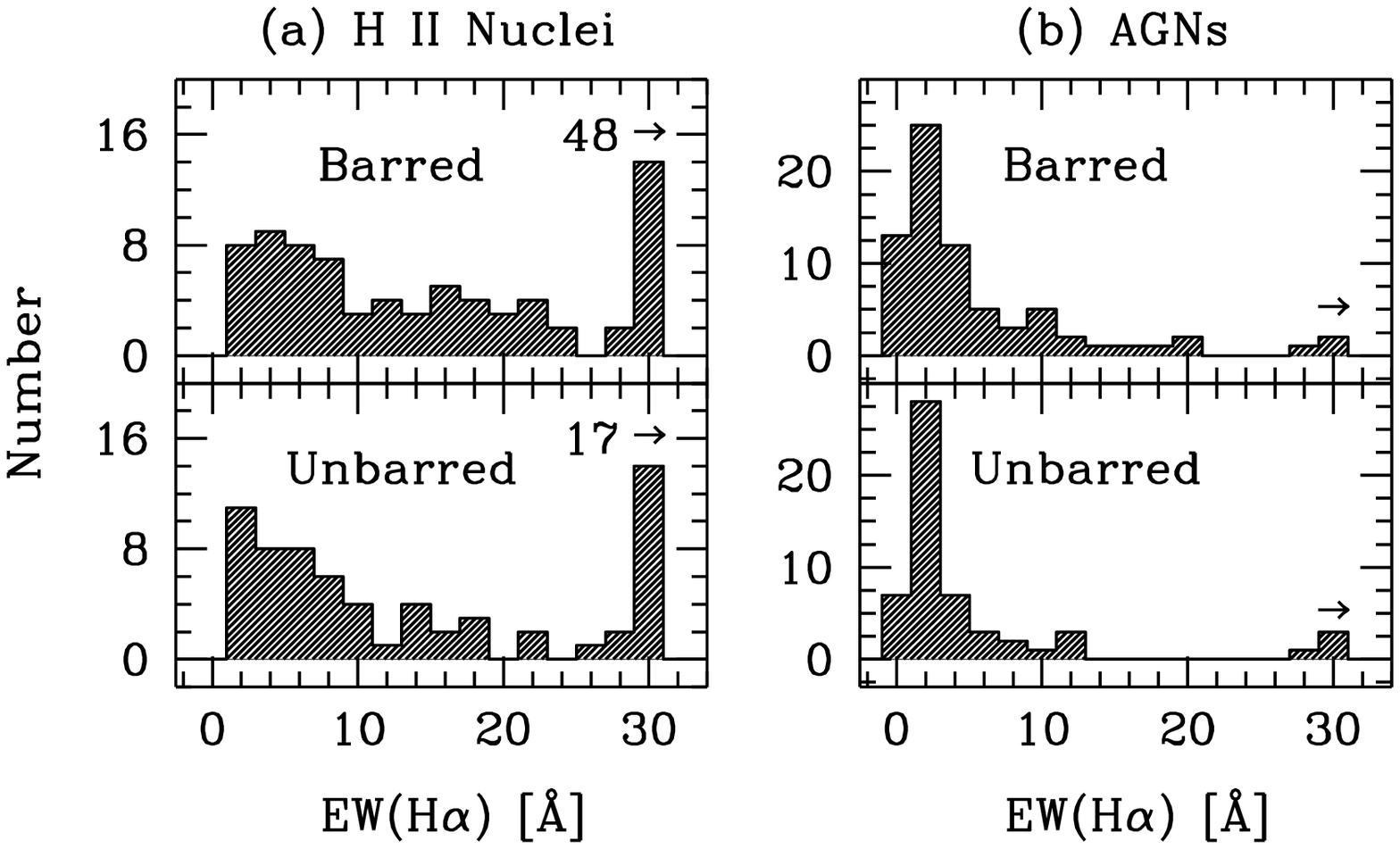}
\caption{}
\end{figure}

\clearpage
\begin{figure}
\figurenum{6}
\plotone{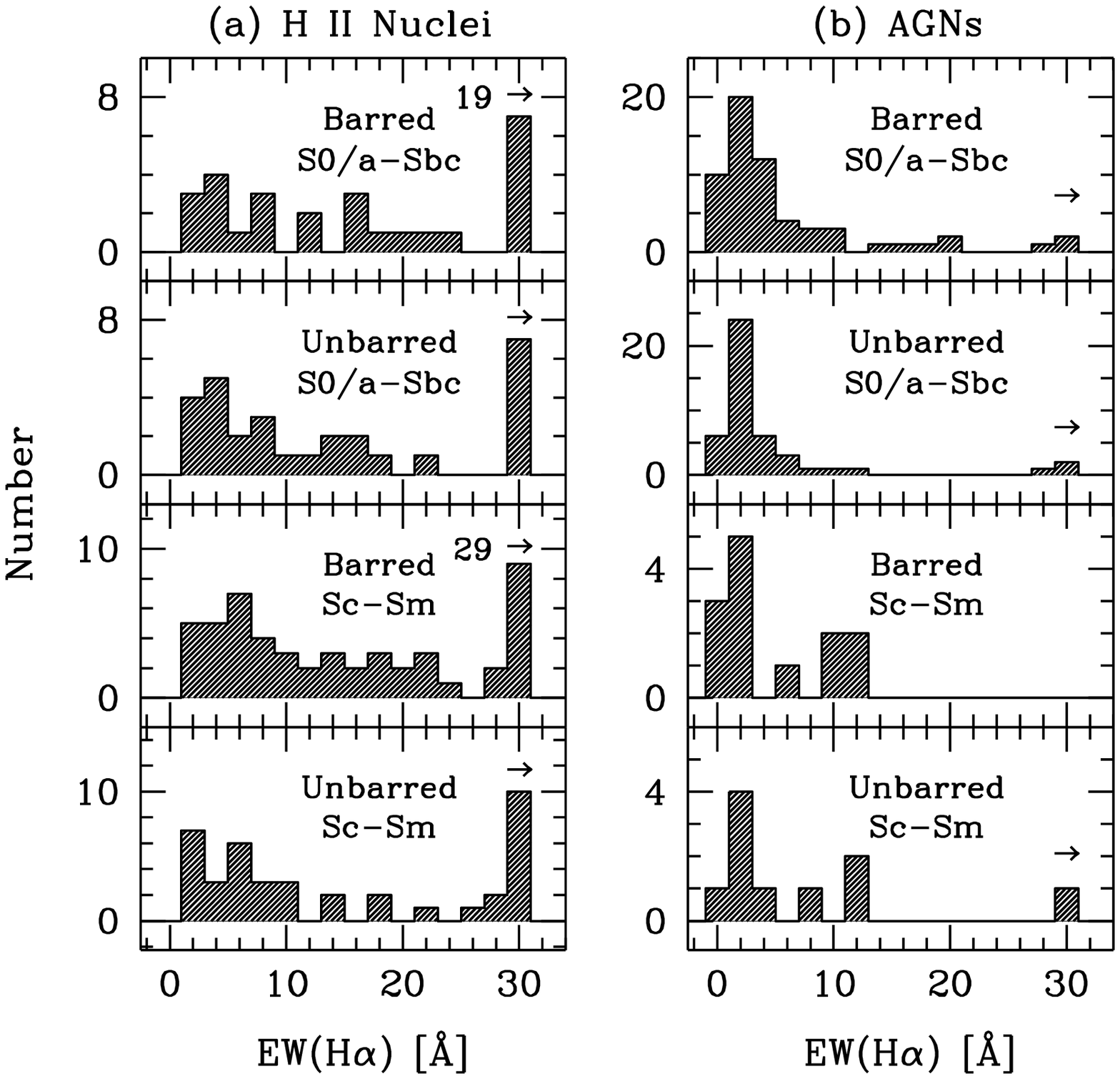}
\caption{}
\end{figure}

\end{document}